# DOWN GOING MUON RATE MONITORING IN THE ANTARES DETECTOR

K. Gracheva[1], M. Anghinolfi[2], V. Kulikovskiy[1,2], E. Shirokov[1], Y. Yakovenko[1]

[1]*Moscow State University, Leninskie Gori,119991 Moscow*
[2]*Dipartimento di Fisica dell'Università e INFN- Via Dodecaneso 33, 16146 Genova*

**Abstract**

Large underwater telescopes have been proposed as a challenging method to measure high energy neutrinos from astrophysical objects. In recent years, The Antares collaboration has designed and realized the first detector of this type in the Mediterranean Sea. Muon tracks produced by the neutrino interaction in the surrounding medium are reconstructed from the arrival time and the number of photo-electrons of the Cherenkov light measured by the Photomultiplier tubes (PMT) array of the detector.

In order to provide sufficient statistics, the events from various periods in the year must be summed together taking care of the various environmental conditions and detector configurations.

In this note we describe effective criteria to group compatible runs based on the effective number of active PMTs in each run.

# 1.INTRODUCTION

The ANTARES detector, located 40 km off-shore from Toulon at 2475m depth, was completed on M ay 2008, making it the largest neutrino telescope in the northern hemisphere[1]. The detector infrastructure has 12 mooring lines holding light sensors designed for the measurement of neutrino induced charged particles based onthe detection of Cherenkov light emitted in water.

The Data Acquisition system (DAQ) is based on t he all-data-to-shore concept. In this mode, all signals from the PMTs that pass a preset threshold (typically 0.3 photo-electron) are digitized and sent to shore where they are processed in real-time by a farm of 250 PCs. The data filter algorithm applied onshore is based on different trigger criteria, including a general purpose muon trigger(3N), a directional trigger (the Galactic Centre trigger, GC), muon triggers based on local coincidences(2T3), a minimum bias trigger for monitoring the data quality, and dedicated triggers for multi-messenger investigations. The filtered data are written to disk in ROOT format by a central data writing process and copied every night to the computer centre in Lyon.

After calibration on position, timing and amplitude, muon tracks are reconstructed using the general causality relation:

$$|t_i - t_j| \leq |x_i - x_j| \times \frac{n}{c} + 20ns$$

where $t_i(x_i)$ refers to the time (PMT position) of hit i, $c$ is the speed of light and n the index of refraction of the sea water. This linear pre-fit is followed by another technique based on the M-estimator and a maximum likelihood method that includes the likelihood of hits from background light[2]. The observed trigger rate is dominated by the background of atmospheric muons and amounts to 5-10 Hz depending on t he trigger conditions.

Once the reconstructed muon tracks are recorded, the following part of the analysis includes the data quality checks to determine the consistency between the various runs collected in different environmental conditions.

This note describes the method we have adopted to link the measured rate of muon tracks (mostly down going) to the detector configuration, trigger conditions, background rate and to define subsets of compatible runs which can be summed to improve statistics.

# 2. THE MUON RATE

The following figure represents as an example the rate of down going reconstructed muons
evaluated for each Antares run in 2009. Naively one should expect that in an ideal detector this rate is kept constant. On the contrary, strong fluctuations can be observed, up to a factor four, according to the different detector configuration including trigger type, environmental conditions and number of active photomultipliers (PMT).

The last one is intuitively the most effective in the muon reconstruction. In particular, due to the bioluminescence activity in the detector environment, even if during a single run the number of working PMTs is constant, not all the data from them are available offshore due to Xoff and HRV. HRV (High Rate Veto) condition occurs when in one single PMT, due to a bioluminescence burst, the single hit rate exceeds a given threshold (typically 500 kHz) and no data are sent to shore. The same occurs for Xoff when the PMT registers an excessive counting rate for a long period overfilling the buffer.

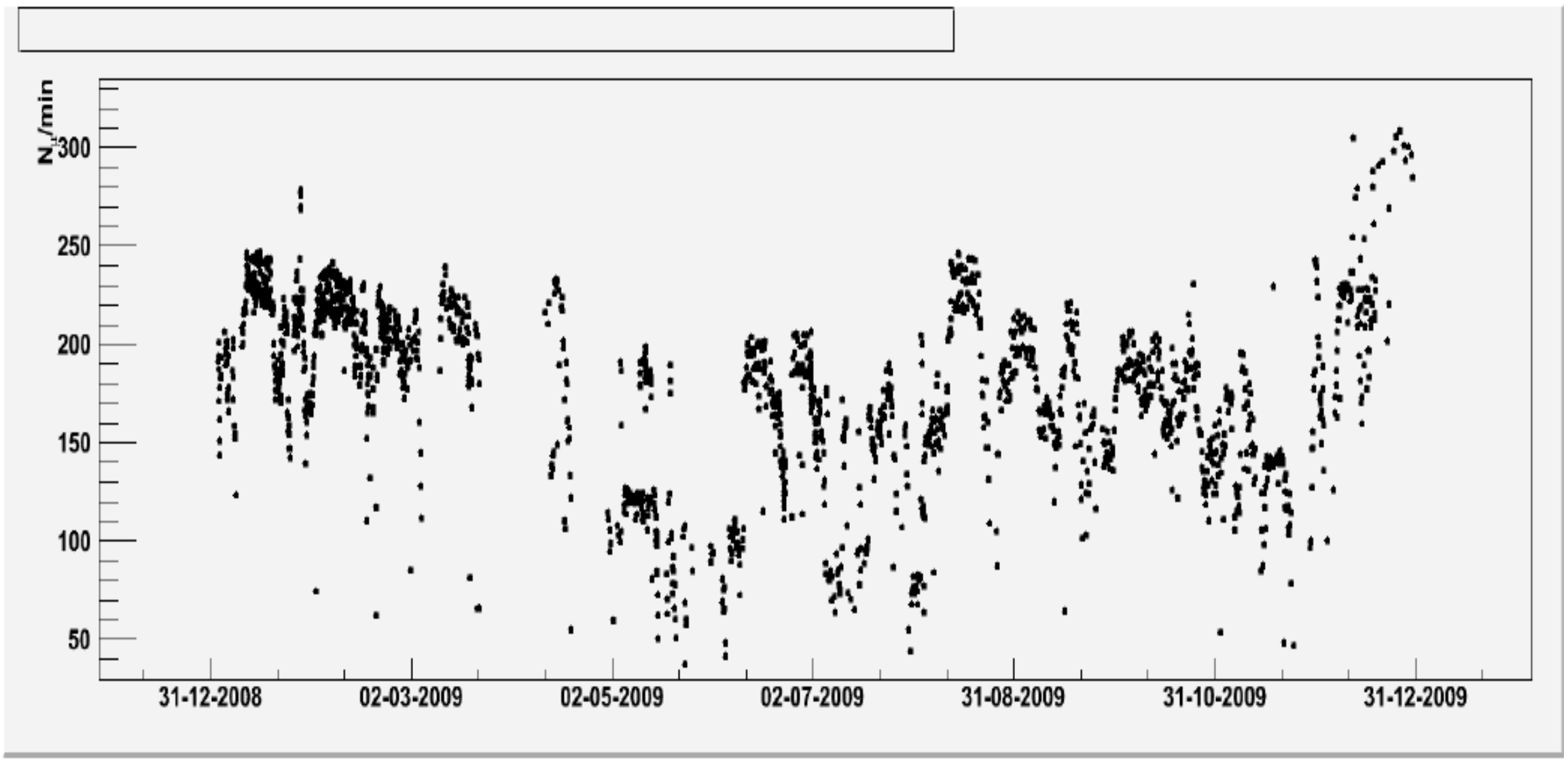

**FIG. 1** - Average muon rate for various runs in 2009.

The information of each i-th PMT, including its total number of hits $h_i$ in one time slice (time slice =0.105 s), is denoted as the i-th frame $F_i$, the total number of frames being the same as the total number of PMTs. A frame is active when its corresponding PMT is active i.e. when $h_i$>1000. The average number of active PMTs $A_{PMT}$ in one run is therefore:

$$A_{PMT}=\frac{\sum_{i=1}^{N_{timeslices}} F^i_{>1000}}{N_{timeslices}}$$

where $F^i_{>1000}$ are frames for which $h_i$>1000. The muon rate expressed in this new variable for a set of runs taken within the same physics trigger but into two detector configurations is reported in fig.2. Here we notice two separate regions corresponding to the two detector configurations with 49 and 54 active sectors respectively. However, differently from fig.1, the muon rate on each region is well aligned to the number of active PMT. To understand this dependence, one can start considering for each run, the distribution of the number of hits $n_{hit}$ used in the reconstruction of the muon tracks where the cut

$$n_{hit} \geq 6 \qquad (1)$$

has been used. From the plot, shown in fig. 3, the number of muon $n_\mu$ reconstructed with $n_{hit}$ number of hits is:

$$n_\mu(n_{hit}) = \alpha e^{-\beta(n_{hit}-6)}$$

If N is the number of active PMT of the run, the total number of reconstructed muons $\mu_N$ is:

$$\mu_N=\sum_{n_{hit}=6}^{\infty} \alpha e^{-\beta(n_{hit}-6)} = \alpha \sum_{x=0}^{\infty} e^{-\beta x} = \frac{\alpha}{1-e^{-\beta}} \quad (2)$$

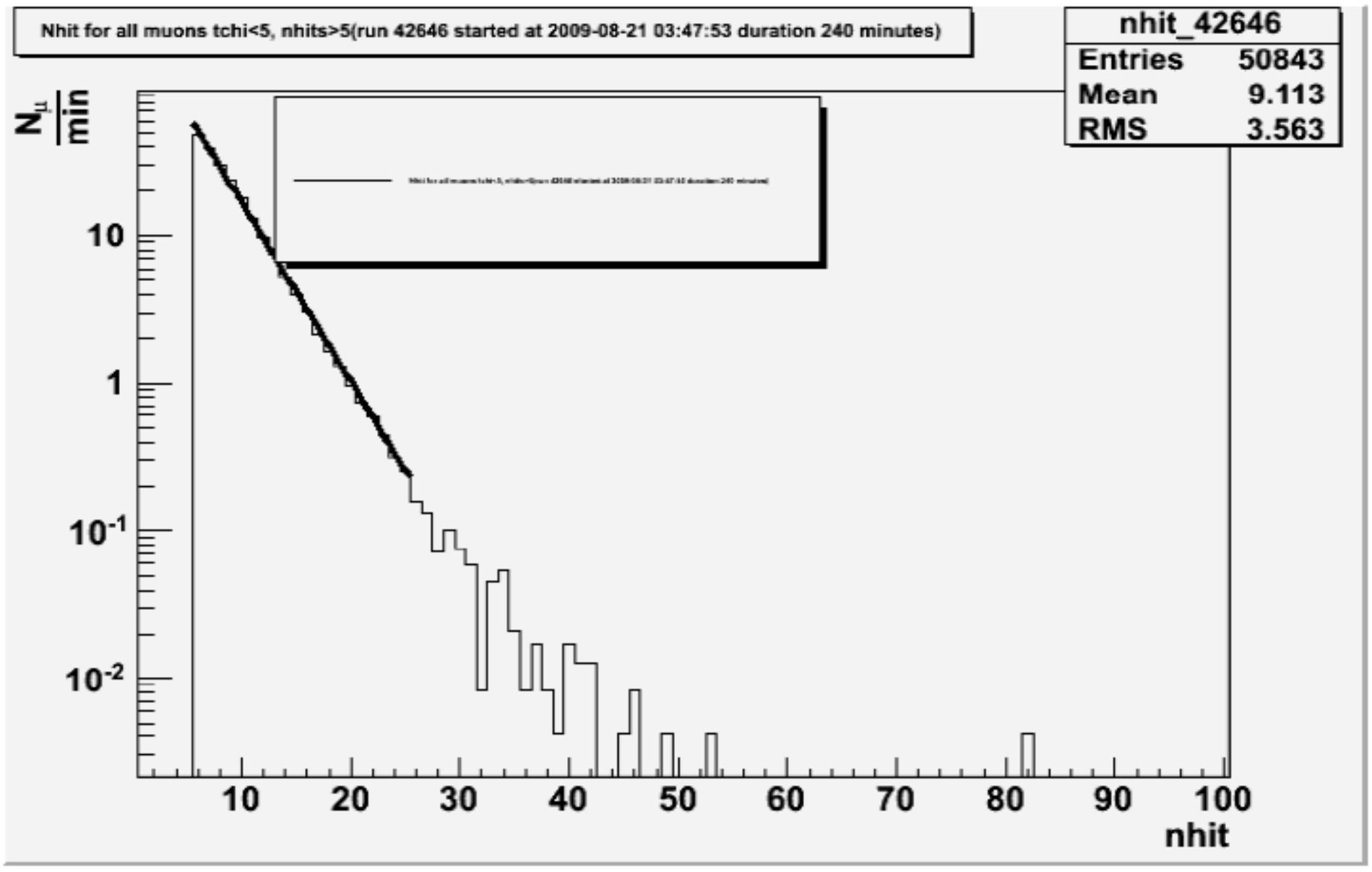

**FIG. 2 -** Average muon rate as a function of the number of active PMT defined in the text. The two groups correspond to two different sector numbers.

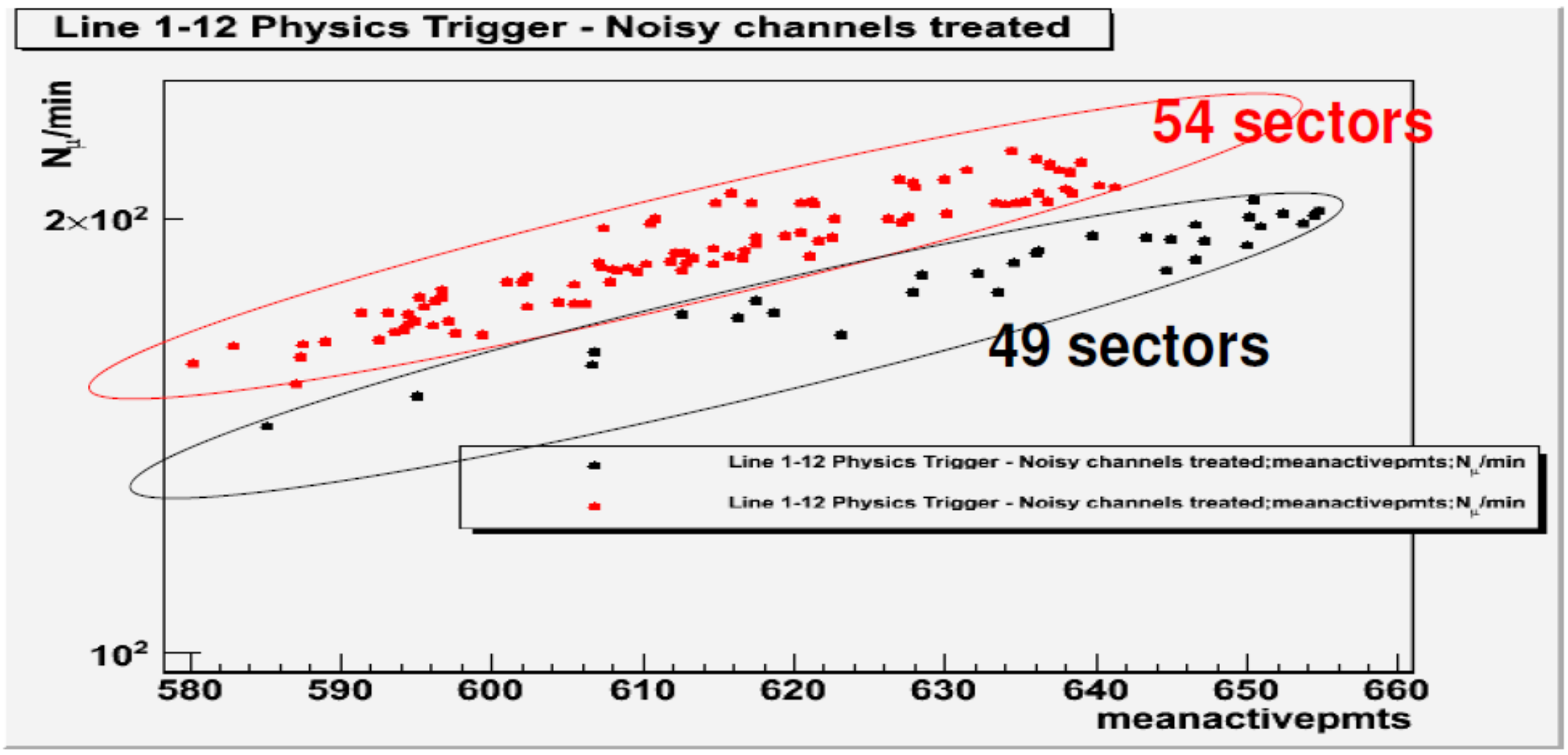

**FIG. 3 -** The distribution of the number of hits used in the reconstruction of a muon track.

Let us assume now that the active number of PTM is decreased by one unit: in that case only some of the $\alpha$events in the first bin will be lost because they do not satisfy condition (1) any more. In particular we expect to lose one event out N/6 for a total of

$$\mu_{N-1} - \mu_N = \alpha \frac{6}{N}$$

Therefore the number of reconstructed tracks for N-1 active PMTs becomes:

$$\mu_{N-1} = \mu_N (1 - \frac{6}{N}(1 - e^{-\beta}))$$

where eq. (2) was used to determine α.
In general, for N-δ active PMTs, we can write

$$\mu_{N-\delta} = \mu_N \left(1 - \frac{6}{N}(1 - e^{-\beta})\right)^{\delta} = \mu_N C^{\delta}$$

where we assume δ<< N and β constant, C= $1 - \frac{6}{N}(1 - e^{-\beta})$

The relation between $\mu_{N-\delta}$ and $\mu_N$ reads that the reconstructed tracks in a given trigger condition depends exponentially from the number of active PMT. In particular, being β fairly constant as seen in fig. 4, we can expect that the muon rate is aligned with respect to the number of active PMT as found in the log scale of fig. 3.

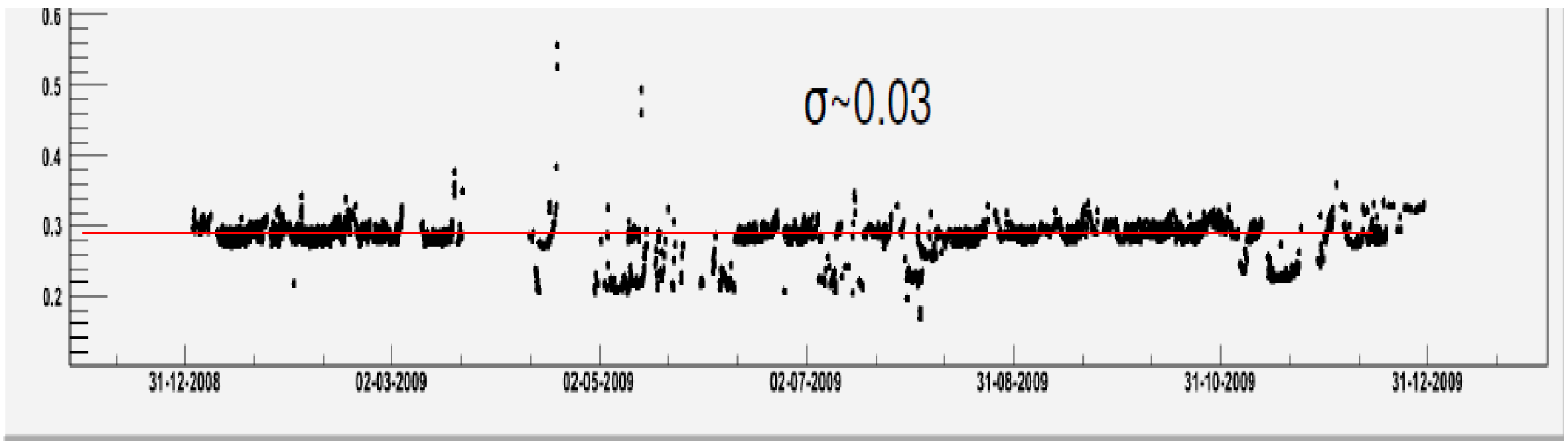

**FIG. 4** - The parameter β in eq. 2 for various runs.

In particular, from the same figure and from fig. 4 we can verify that, in the configuration of 54 active sector, the expected muon rate with 590 active PMTs μ(590) is related to the muon flux with 640 active PMTs,μ(640), by

$$\mu(590) \cong \mu(640)\left(1 - \frac{6}{640}(1 - e^{-0.3})\right)^{50} \cong 200\ 0.9976^{50} \cong 177$$

as seen in the figure.

**3. THE DATA SELECTION**

While the two regions in fig. 2 suggest that the muon rate with respect to the number of active PMTs clusters in different lines according to different detector configurations, we now investigate the dependence from the trigger conditions, given a fixed number of sectors. As an example, in fig. 5, we report the muon rate for different triggers in the same geometrical configuration of the detector. The trigger can be varied according to bioliminescence conditions.

For low activity, the 3N+ 2T3+GC is currently used where GC is a directional trigger which enhances detection efficiency in the direction of the centre of the galaxy while for increasing activity the 3N+ 2T3 and then the 3N only are used. The shape of the cluster in fig.5 suggests that trigger conditions weakly affects the muon rate though a larger scatter of the poi nts is observed i n the presence of t he GC condition. This is easlily understood from the top of fig.6 where the muon rate is plotted as a function of the time during a week. The clear modulation of one sideral day period is an effect of the GC trigger which enhance detection efficiency by 15% each time the galactic centre is visible for Antares.

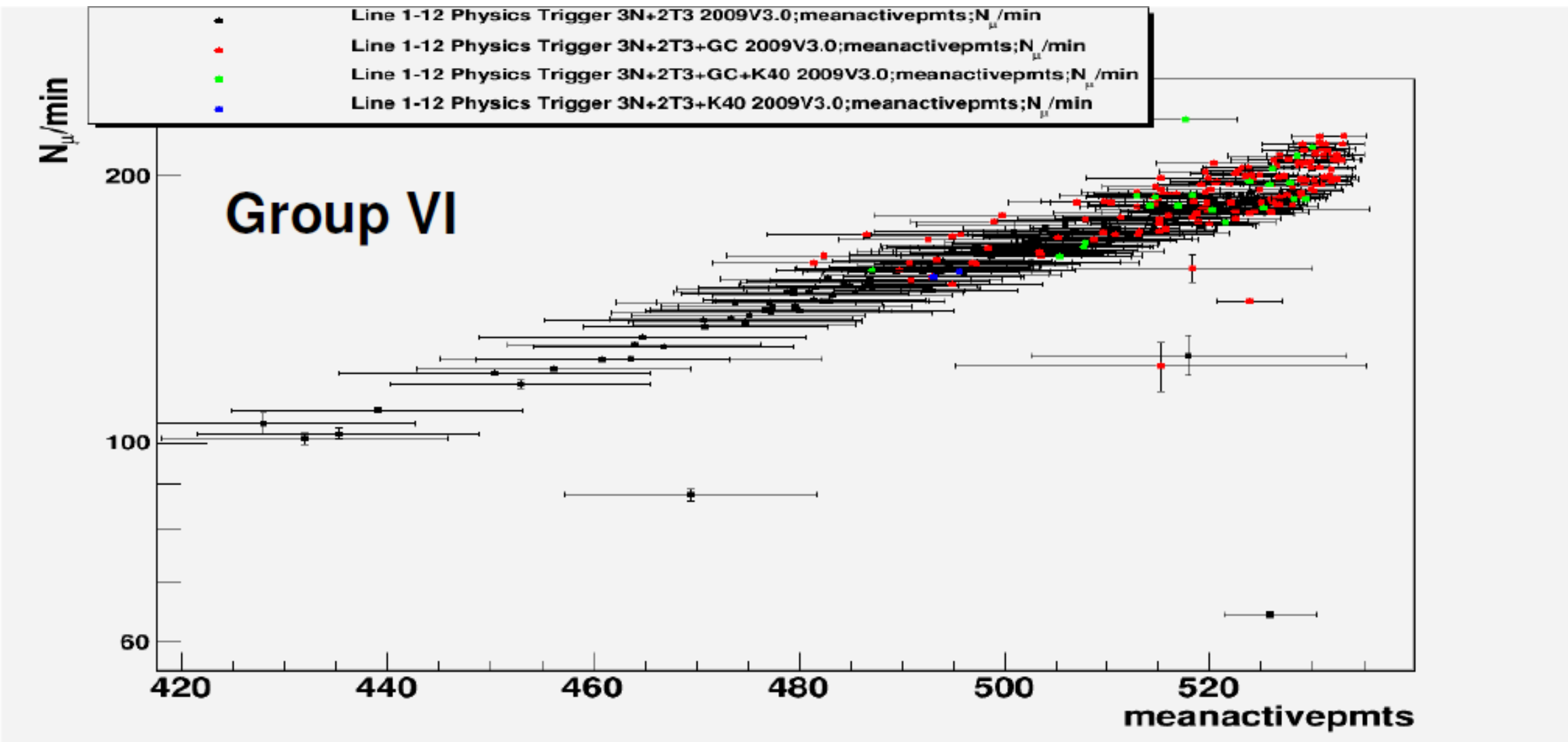

**FIG. 5** - The muon rate for different triggers in the same geometrical configuration of the detector.

From the effects we have described we can deduce an effective procedure to select the runs which are compatible and which can be summed together:

- calculate the number of active PMT $A_{PMT}$ in the run,
- plot the reconstructed muon rate/minute as a function of $A_{PMT}$ to check the alignment,
- select different clusters according to the number of active sectors in the detector as in fig. 2,
- for each cluster, determine the distance of each point from the interpolation line,
- choose a cut on this distance, typically 4σ of the distribution.

An example of this procedure is shown in fig. 7 where the muon rate for the runs collected with both the 3N+2T3+GC and 3N+2T3 triggers (same as fig. 5) is plotted as a function of $A_{PMT}$. The distribution of the distance of each point from the average line is plotted on the right: only the runs corresponding to the central peak in the distribution were accepted as compatible.

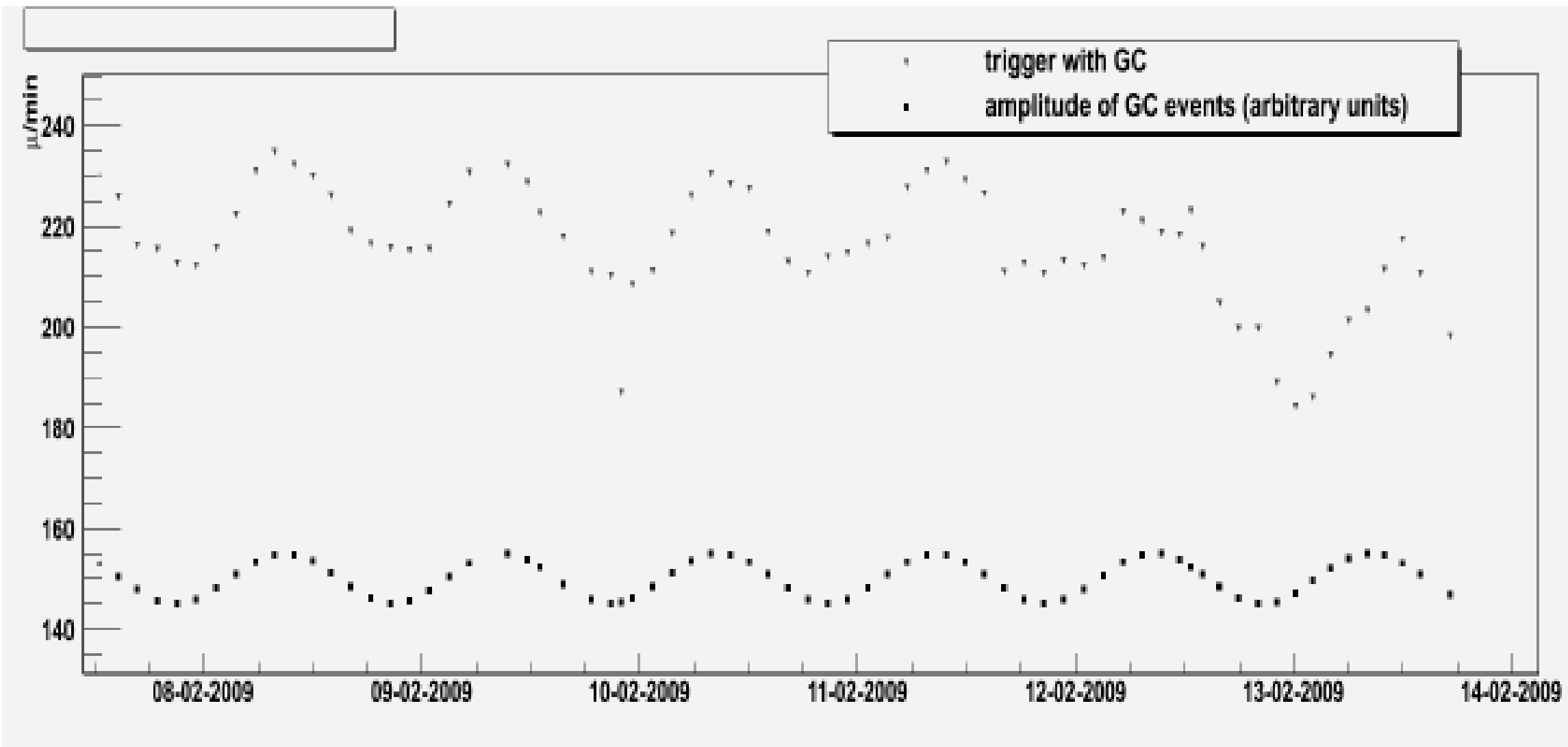

**FIG. 6** - The muon rate (top) for the runs collected activating the Galactic Centertrigger. The sideral modulation is clearly evident.

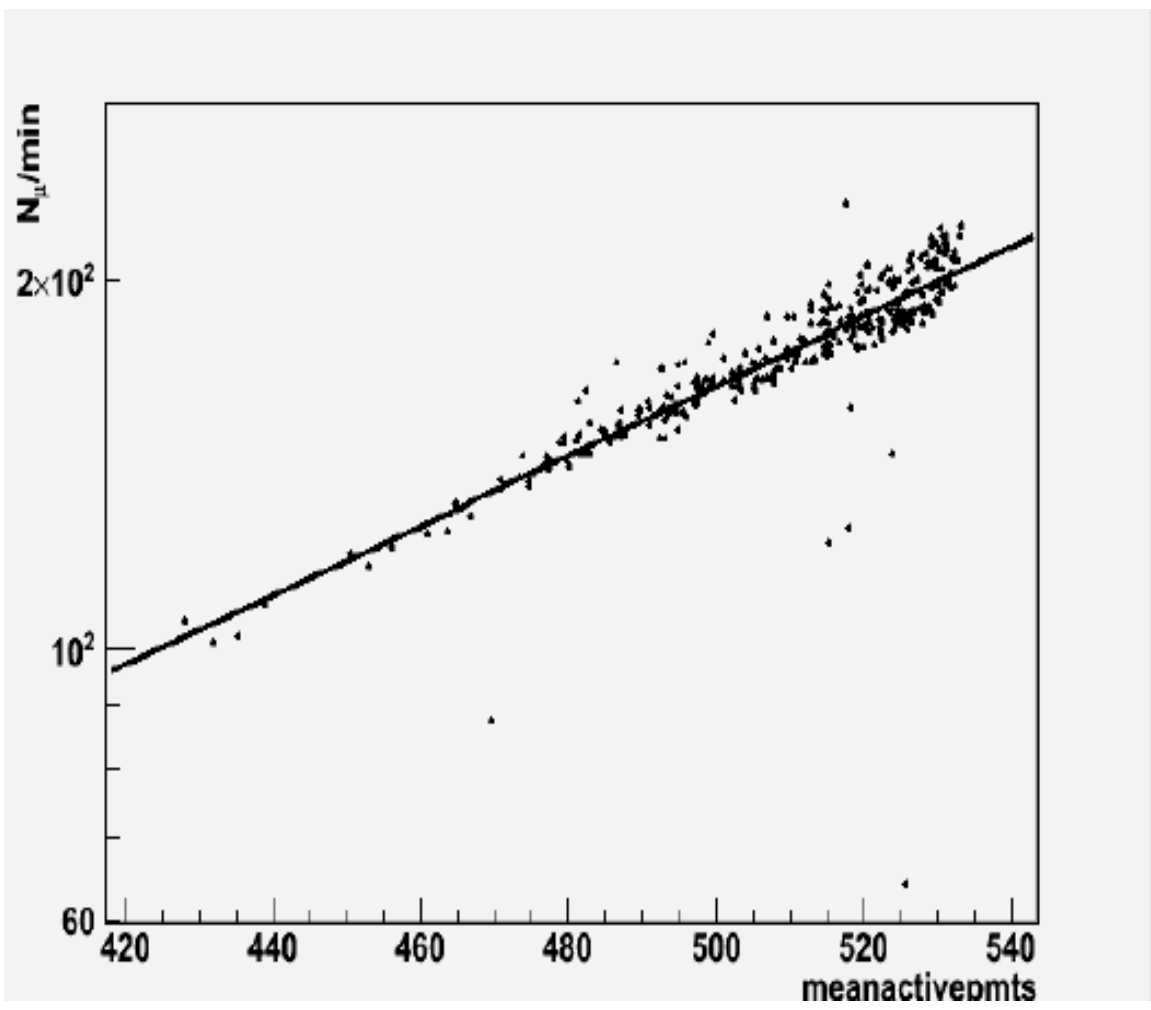

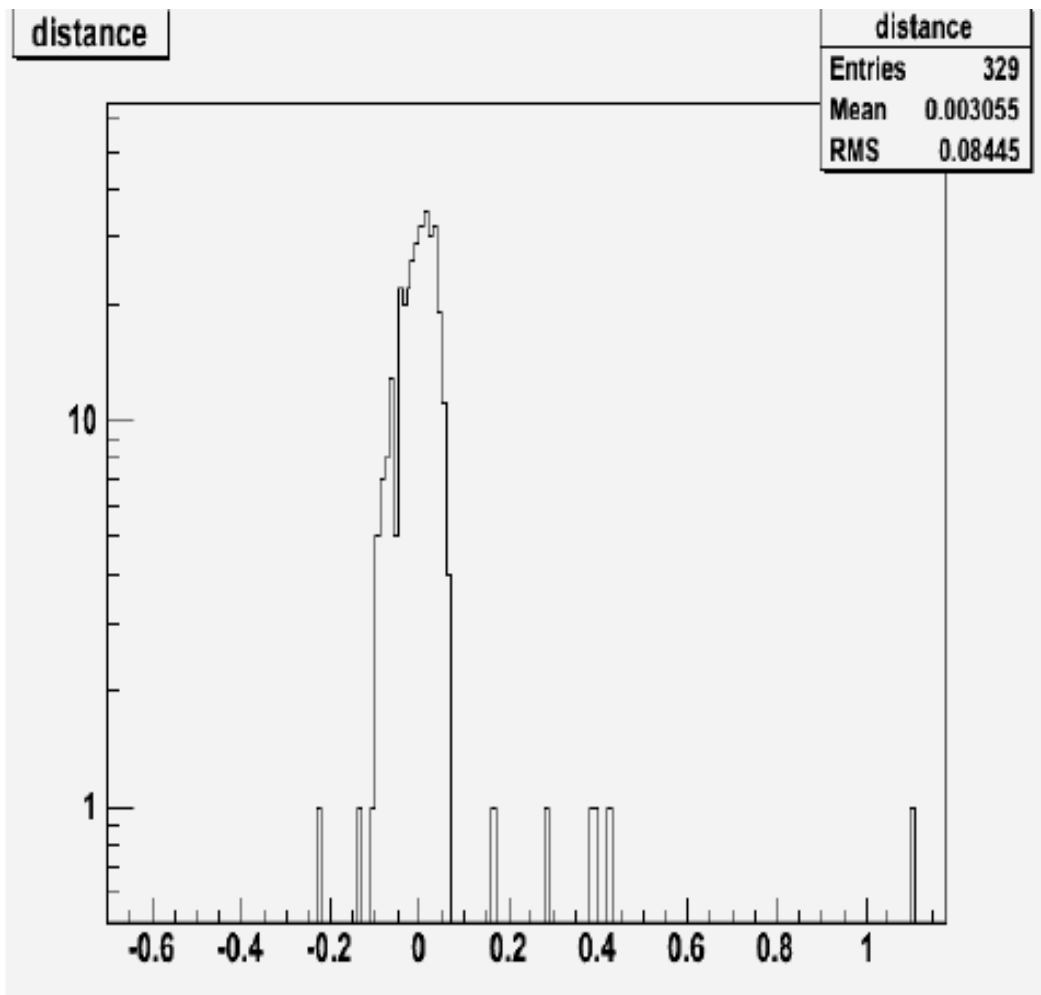

**FIG. 7 -** The muon rate as a function of the active PMTs (left) and the histogram of the distance of the points from the line of interpolation (right).

The following table summarizes the selection of the run taken in2009: here up to 9 groups are found according to the different number of active lines and sectors (last three columns). The number of runs which have been excluded according to the described procedure is found in the column 'Bad runs' and accounts up to a maximum of 5.7%.

**TABLE 1** - Classification of the runs in Antares during 2009.

| № | runs | Period | Runs total | Bad runs | % | Sectors | lines | Comments |
|---|---|---|---|---|---|---|---|---|
| I | 38347-38357 | 02.01.2009-07.01.2009 | 30 | 0 | 0 | 54 | 12 | |
| II | 38363-39345 | 08.01.2009-26.02.2009 | 479 | 14 | 2.3 | 49 | 11 | |
| III | 39349-39589 | 27.02.2009-12.03.2009 | 52 | 3 | 5.7 | 48 | 11 | |
| IV | 39590-41677 | 12.03.2009-02.07.2009 | 443 | 21 | 4.7 | 46 | 10 | High bioactivity |
| V | 41679-42652 | 02.07.2009-21.08.2009 | 274 | 8 | 2,9 | 43 | 9 | HV tuning |
| VI | 42658-44109 | 22.08.2009-27.10.2009 | 329 | 6 | 1.8 | 41 | 9 | |
| VII | 44115-44326 | 28.10.2009-06.11.2009 | 49 | 1 | 2 | 40 | 8 | |
| VIII | 44328-44461 | 07.11.2009-12.11.2009 | 37 | 0 | 0 | 45 | 9 | 10th line adjustment |
| IX | 44472-45503 | 14.11.2009-31.12.2009 | 182 | 6 | 3.3 | 49 | 10 | 12th line adjustment |

**CONCLUSIONS**

The measured muon rate on an ideal underwater detector should be almost constant on time. Different environmental conditions, detector and trigger configurations no longer maintain this expectation and criteria to find the compatibility of different runs must be provided.

In this paper we describe an effective method based on the number of active PMT which can be used to select compatible runs. As an example, the method was applied to the 2009 data where 9 different periods where runs can be summed together were found.